%% file: The_Complexity_of_Contracting_Planar_Tensor_Networks.tex
\documentclass[a4paper,UKenglish,cleveref, autoref, thm-restate]{lipics-v2021}

\usepackage[ruled]{algorithm2e}
\usepackage{changepage}
\usepackage{array}
\usepackage{multirow}
\usepackage{multicol}
\usepackage{makecell}
\usepackage{caption}
\usepackage{subcaption}
\usepackage{bm}

\title{The Complexity of Contracting Planar Tensor Networks} 

\author{Ying Liu \footnote{Corresponding author}}{State Key Laboratory of Computer Science, China \and Institute of Software Chinese Academy of Sciences, China \and University of Chinese Academy of Sciences, China }{liuy@ios.ac.cn}{}{Supported by NSFC61932002 and NSFC 61872340.}

\authorrunning{\ } 

\Copyright{\ } 


\keywords{complexity, finite element graphs, exponential time, planar graphs, tensor network contraction, \#ETH, \#3SAT} 

\category{} 

\relatedversion{} 

\acknowledgements{The authors are very grateful to Prof. Jinyi Cai and Prof. Mingji Xia for their beneficial guidance and advise.}

\EventEditors{John Q. Open and Joan R. Access}
\EventNoEds{2}
\EventLongTitle{42nd Conference on Very Important Topics (CVIT 2016)}
\EventShortTitle{CVIT 2016}
\EventAcronym{CVIT}
\EventYear{2016}
\EventDate{December 24--27, 2016}
\EventLocation{Little Whinging, United Kingdom}
\EventLogo{}
\SeriesVolume{42}
\ArticleNo{23}
\begin{document}
	\nolinenumbers
	\maketitle
	
	\begin{abstract}
		\quad\ Tensor networks have been an important concept and technique in many research areas, such as quantum computation and machine learning. We study the exponential complexity of contracting tensor networks on two special graph structures: planar graphs and finite element graphs. 
		
		We prove that any finite element graph has a $O(d\sqrt{\max\{\Delta,d\}N})$ size edge separator. Furthermore, we develop a $2^{O(d\sqrt{\max\{\Delta,d\}N})}$ time algorithm to contracting a tensor network consisting of $N$ Boolean tensors, whose underlying graph is a finite element graph with maximum degree $\Delta$ and has no face with more than $d$ boundary edges in the planar skeleton, based on the $2^{O(\sqrt{\Delta N})}$ time algorithm \cite{fastcounting} for planar Boolean tensor network contractions.
		
		We use two methods to accelerate the exponential algorithms by transferring high-dimensional tensors to low-dimensional tensors. 
		We put up a $O(k)$ size planar gadget for any Boolean symmetric tensor of dimension $k$, where the gadget only consists of Boolean tensors with dimension no more than $5$. 
		Another method is decomposing any tensor into a series of vectors (unary functions), according to its \emph{CP decomposition} \cite{tensor-rank}. 
		
		We also prove the sub-exponential time lower bound for contracting tensor networks under the counting \emph{Exponential Time Hypothesis} (\#ETH) holds.
	\end{abstract}

	\input{introduction.tex}

	\input{preliminaries.tex}

	\input{algorithm.tex}

	\input{lowerbound.tex}

	\input{conclusion.tex}

	\bibliography{ref.bib}
	
	\newpage
	\appendix

\input{appendix.tex}

\end{document}

%% file: introduction.tex
\section{Introduction}
	\quad\ Tensor network states provide an import analytic framework for high dimensional data structures among a variety of scientific disciplines, such as deep convolutional arithmetic circuits in machine learning \cite{LYCS2019}, partition functions in classical statistical mechanics \cite{EV2015}, quantum circuits in quantum theory \cite{MS2008, CZCHNS2018, VBNHRBM2019, GK2021} and more. The underlying idea of tensor networks is to use the interconnection among low-dimensional tensors to represent the complex entangled data structures so people can easily manipulate them.
	The \emph{vaule} of a tensor network, a scalar quantity obtained by contracting all tensors, reflects some critical characteristics of the corresponding data structure. For example, the amplitude of a quantum circuit \cite{CZCHNS2018}. Therefore, the complexity of tensor network computation or tensor network contraction is a worthy research topic.

	The order of edges in the contraction sequence would not affect the value of a tensor network, but it decides the total time cost of the contraction process.
	So choosing an optimal contraction sequence is the key to efficiently computing the value of a tensor network. 
	Markov and Shi \cite{MS2008} demonstrated that the optimal contraction sequence is decided by the minimum-width tree decomposition of the line graph of the underlying graph. Thus, the total contraction time is closely related to the treewidth of the line graph. 
	Finding the tree decomposition with minimum width is an NP-hard problem, and searching for an optimal contraction sequence is also hard.
	People prefer constructing a sequence that makes the number of high arithmetic-intensity contractions small. A common way is searching for an appropriate cut or separator hierarchy to construct the contraction sequence \cite{SHHMSS2016, AHSS2017, fastcounting, GK2021}, which also involves the aspect of \emph{graph partition}. 

	Researchers usually preprocess the tensor network to construct the contraction sequence more efficiently. A universal method to simplify a tensor network is called \emph{tensor slicing} or \emph{tensor decomposition}, which decomposes a tensor into some lower dimensional tensors. A series of tensor decomposition formats, presented in \cite{tensor-rank} and \cite{CLOPZM2016}, have been introduced and exploited in practice in several works, for example, the SVD \cite{GK2021} and Tensor-Train decomposition \cite{KCKXM2022} for the tensor network simplification. 

	A tensor of some dimension $k$ is actually a function of arity $k$.
	The parameterized set of functions $\mathcal{F}$ is a decisive factor of the complexity of a tensor network computation problem, where each tensor belongs to $\mathcal{F}$ in the input. 
	Tensor network contraction is exactly the class \emph{Holant Problem} \cite{HolantProblem}, and the computational complexity of the Holant problem has been widely studied in past years. 
	Cai et al. presented a series of tractable conditions, which $\mathcal{F}$ satisfies, such that the corresponding Holant problem is tractable in polynomial time; otherwise it is \#P-hard  \cite{complexHolant*,complexHolantc,nonnegativeHolant,realHolant}. Dell et al. \cite{ETHKSAT, SETH} imported the counting version of \emph{Exponential Time Hypothesis} to study the fine-grained complexity classification of some counting problems \cite{ETHpermTutte, ETHquantumGH, blockinterpolation}, which are tensor network contraction problems with special parameterized function sets. They demonstrate that such a problem is polynomial-time solvable if the parameterized set $\mathcal{F}$ satisfies the given condition; otherwise the problem can not be computed in sub-exponential time when the counting Exponential Time Hypothesis (\#ETH) holds. 
	Besides, the excellent performance of \#ETH in proving sub-exponential lower bounds of tensor network contraction problems also has been confirmed in \cite{ETHCSP,exponentialdichotomyofCSP}.
		
	\subsection{Main results}
	
	\quad\ We focus on the algorithms and computational lower bounds of tensor network contraction problems on two special graph structures: planar graphs and finite element graphs.
	Main results are presented in Table \ref{table1}.
	
	\begin{table}[ht]
		\centering	
		\renewcommand\arraystretch{1.5}
		\caption{Main results in this article.}
		\label{table1}
		\begin{subtable}[t]{\linewidth}
			\centering
			\caption{The upper bound of the size of edge separator in a graph with $N$ vertices}
				\begin{tabular}{|c|c|}
					\hline
					\textbf{Graph structure} & \textbf{Edge separator size} \\
					\hline
					\textbf{planar graph} & $O(\sqrt{\Delta N})$ \cite{edgeseparator}\\
					\hline
					\textbf{finite element graph} & $O(d\sqrt{\max\{\Delta,d\}N})$ (Theorem \ref{thm1})\\
					\hline
				\end{tabular}
		\end{subtable}
		\begin{subtable}[t]{\linewidth}
			\vspace{0.5cm}
			\centering	
			\caption{The complexity of contracting a tensor network parameterized by the set $\mathcal{F}$.}
			\begin{tabular}{|c|c|c|c|}
				\hline
				\textbf{Graph structure} & $\bm{\mathcal{F}}$ & \textbf{upper bound} & \textbf{lower bound} \\
				\hline
				\multirow{3}{*}{\textbf{planar graph}} & \makecell{a set of \\Boolean functions} & $2^{O(\sqrt{\Delta N})}$ \cite{fastcounting}  &\multirow{3}{*}{$2^{o(\sqrt{N})}$ (Theorem \ref{thm7})}\\
				\cline{2-3}
				&\makecell{a set of \\ symmetric \\  Boolean functions}  & $2^{O(\sqrt{N})}$ (Theorem \ref{thm4})  &  \\
				\cline{2-3}
				&a finite set & $R^{O(\sqrt{N})}$ (Theorem \ref{thm3}) &  \\
				\hline
				\multirow{3}{*}{\makecell{\textbf{finite elements}\\ \textbf{graph}}} & \makecell{a set of \\Boolean functions} & $2^{O(d\sqrt{\max\{\Delta,d\}N})}$ (Theorem \ref{thm2})  & \multirow{3}{*}{$2^{o(\sqrt{N})}$ (Theorem \ref{thm8})}\\
				\cline{2-3}
				& \makecell{a set of \\symmetric \\  Boolean functions} & $2^{O(\sqrt{d^3N})}$ (Theorem \ref{thm6})&  \\
				\cline{2-3}
				& a finite set & $R^{O(d\sqrt{N})}$ (Theorem \ref{thm5})&  \\
				\hline
			\end{tabular}
		\end{subtable}
	\vspace{-0.5cm}
	\end{table}

	Based on the classical planar edge separator theorem \cite{edgeseparator}, we prove that we can find a $O(d\sqrt{\mathrm{max}\{\Delta,d\}N})$ size edge separator for a finite element graph where no face has more than $d$ boundary edges and each vertex has no more than $\Delta$ incident edges. 

	Kourtis et al. \cite{fastcounting} used the planar edge separator theorem \cite{edgeseparator} to find the contraction sequence and put up a $2^{O(\sqrt{\Delta N})}$ time algorithm for planar Boolean tensor network contractions.
	Inspired by it, we develop the $2^{O(d\sqrt{\max\{\Delta,d\}N})}$ time algorithm for Boolean tensor networks on finite elements graphs.

	We preprocess the input tensor network to accelerate the above two algorithms.
	We put up an $O(k)$-size planar gadget for any Boolean symmetric function of arity $k$. Only functions of small arity are applied in such a gadget. Based on this design, we preprocess any tensor network with only Boolean symmetric functions by transferring high dimensional tensors to a series of low dimensional tensors. We eliminate the factor $\Delta$ in the time cost of the contraction algorithms.
	Another preprocessing uses \emph{CP decomposition} \cite{CPdecomposition} to decompose a tensor into a sum of the products of some vectors. 
	We use this tool crossing the node separator hierarchy to provide two divide-and-conquer algorithms for tensor network contraction problems on planar and finite elements graphs, respectively. Let $R$ denote the maximum rank of the functions in $\mathcal{F}$. The two algorithms cost $R^{O(\sqrt{N})}$ and $R^{O(d\sqrt{N})}$ time, respectively.

	Naturally, we consider the computational lower bound of tensor network contractions.
	We present $2^{o(N)}$ time lower bound for tensor networks defined on the set $\{=_2,=_3,\neq_2,OR_3\}$, even restricted on planar or finite element graphs.

%% file: preliminaries.tex
\section{Preliminaries}

\subsection{Definitions and notations}

	\quad\ Let $\mathbb{N}$ or $\mathbb{C}$ denote the set of natural numbers or algebraic complex numbers. $[q]$ denotes the finite domain $\{1,2,...,q\}$ for some positive integer $q$. If $q$=2, then $[q]$ is called the Boolean domain where any variable is assigned $0$ or $1$. $\bar{x}=1-x$ denotes the negation for a Boolean variable $x$. 
	
	A function $F$ of some \emph{arity} $k\in\mathbb{N}$ defined on the domain $[q]$ maps $[q]^k$ to $\mathbb{C}$. 
	If $k=1$ or $2$, then $F$ is a \emph{unary} or \emph{binary} function respectively. 
	A function $F$ of arity $k$ is \emph{symmetric} if $F(x_1,x_2,...,x_k)=F(x_{\pi(1)},x_{\pi(2)},...,x_{\pi(k)})$ for any input $x_1,x_2,...,x_k\in[q]$ under  any permutation $\pi: [k]\to[k]$. 
	Any Boolean symmetric function of arity $k$ can be expressed as $[f_0,f_1,f_2,...,f_k]$ where $f_i$ is the value when the assignment of variables has Hamming weight $i\in\{0,1,...,k\}$. 
	For example, the binary equality function or disequality function can be written as $(=_2)=[1,0,1]$ or $(\neq_2)=[0,1,0]$, respectively.
	A function is also called a \emph{signature} or \emph{constraint}.
	
	An undirected graph $G$ is a tuple $(V,E)$, where $V$ is the vertex set and $E\subseteq \{(u,v)|u,v\in V(G)\}$ is the edge set. 
	$N_G(v)$ (or $N(v)$) denotes the set of edges incident to $v$ in $G$. 
	And $d_v=|N(v)|$ denotes the \emph{degree} of $v$.
	A loop is counted twice, and a $k$-multiple edge is counted $k$ times when computing $d_v$.
	$\Delta=\max\{d_v:v\in V\}$ denotes the maximum degree of $G$. 
	If $\Delta$ is a constant, then $G$ is a bounded degree graph. 
	An edge is called a $bridge$, i.e., a cut-edge, whose deletion increases the number of connected components in the graph.
	A set $S$ of vertices or edges is called a node or edge \emph{separator} if $G$ is partitioned into two disconnected components $A$ and $B$ after removing $S$. $S$ is called a \emph{balanced} separator if $A$ and $B$ both have no more $\frac 2 3|V(G)|$ vertices.
	  
	A graph is planar if it has a planar embedding, i.e., it can be drawn on the plane in such a way that edges intersect only at the endpoints. 
	A balanced node or edge separator for a planar graph can always be found in linear time.
	\begin{lemma}[\cite{nodeseparator}]
		Let $G$ be a planar graph with $N$ vertices.
		A balanced node separator $C\subseteq V(G)$ can be found in $O(N)$ time, such that $|C|=2\sqrt{2N}$.
		\label{planar_node_separator}
	\end{lemma}
	\begin{lemma}[\cite{edgeseparator}]
		Let $G$ be a planar graph with $N$ vertices and maximum degree $\Delta$.
		A balanced edge separator $C\subseteq E(G)$ can be found in $O(N)$ time, such that $|C|=3\sqrt{2\Delta N}$.
		\label{planar_edge_separator}
	\end{lemma}
	
	Consider a class of graphs that are ``almost'' planar.
	A canonical example is \emph{finite element graphs}.
	A finite element graph $G$ is formed from a \emph{planar embedding} $G^*$ of a planar graph by adding all possible diagonals to each face which has more than $3$ boundary edges \footnote{A vertex $v$ would be treated as $k$ different vertices if $k$ incident edges of $v$ are the boundary edges of a face when adding all possible diagonals.}. $G^*$ is called the \emph{skeleton} of $G$, and each faces in $G^*$ is an \emph{element} of $G$. According to the definition, $G$ can be drawn on the planar with crossings only appearing inside the elements.
	\begin{lemma}[\cite{nodeseparator}]
		Let $G$ be a finite element graph with $N$ vertices.
		Suppose each element of $G$ has no more than $d$ boundary edges.
		A balanced node separator $C\subseteq V(G)$ with $|C|=4\lfloor\frac d 2\rfloor \sqrt{N}$ can be found in polynomial time.
		\label{separator_for_finite_element_graph}
	\end{lemma}
	
%
%
%
%
%
%
%
%
	
	A $k$-dimensional \emph{tensor} $F\in\mathbb(C)^{q_1,q_2,...,q_k}$ is also a function of arity $k$ with variables $x_1\in[q_1],x_2\in[q_2],...,x_k\in[q_k]$, for some integers $q_1,q_2,...,q_k$. 
	Let $\mathcal{F}$ denote a set of functions defined on the finite domain $[q]$ for some integer $q$.
	A tensor network defined on $\mathcal{F}$ is a signature grid $(G,\pi)$, where $G(V,E\cup X)$ with two disjoint sets of edges is a graph and $\pi$ maps every vertex $v\in V$ to a function $F_v\in\mathcal{F}$ together with a linear order to $N(v)$.
	$X$ denotes the set of dangling edges with one endpoint in $V$ and the other dangling. 
	The tensor network $(G,\pi)$ defines an $|X|$-dimensional tensor (or a $|X|$-arity function):
	
	\begin{equation}
		\Gamma(y_1,y_2,...,y_k)=\sum_{\sigma: E\to [q]} \prod_{v\in V} F_v(\hat{\sigma}|_{N(v)}),
		\nonumber
	\end{equation}
	where $k=|X|$, $(y_1,y_2,...,y_k)\in[q]^k$ is an assignment to $X$, and $\sigma|_{N(v)}$ is the extension of $\sigma$ on $N(v)$ by $(y_1,y_2,...,y_k)$. 
	We called the tensor network $(G,\pi)$ a \emph{gadget} or an $\mathcal{F}$-gate with the signature $\Gamma$.
	
	If $X=\emptyset$, $(G,\pi)$ is an input of a tensor network contraction problem or a Holant problem parameterized by $\mathcal{F}$. 
	
	\begin{definition}
		Let $\mathcal{F}$ be a set of functions on the domain $[q]$.
		A tensor network contraction problem defined on $\mathcal{F}$, denoted by \#$\mathcal{F}$, is defined as
		
		Input: $(G,\pi)$.
		\begin{equation}\mathit{Output: }
			Z(G)=\sum_{\sigma :E\to [q]} \prod_{v\in V} F_v(\sigma|_{N(v)}).
			\nonumber
		\end{equation} 
	\end{definition}
	The mapping $\pi$ is usually omitted, and $G$ denotes the grid $(G,\pi)$ when the context is clear. 
	The problem $pl$-$\#\mathcal{F}$ denotes the sub-problem of \#$\mathcal{F}$, where all instances are restricted to be planar.
	
	A canonical example is counting Boolean 3-Satisfiability (\#$3$SAT), which is the problem of counting the number of satisfying assignments to a given Boolean 3-CNF. 
	
\begin{figure}[ht]
		\centering
	\includegraphics[scale=0.2]{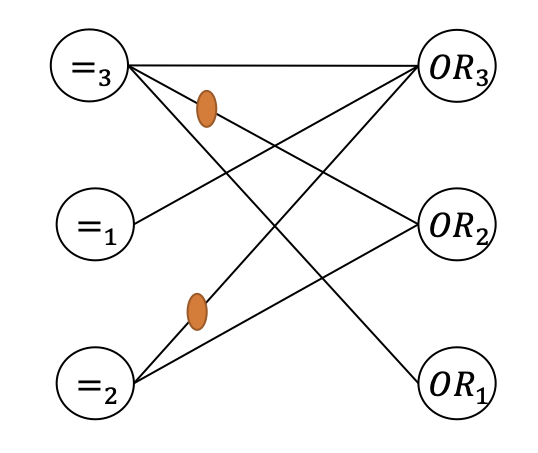}   
	\caption{A tensor network representation for a 3CNF formula $\phi=(x_1\lor x_2\lor \bar{x_3})\land (\bar{x_1}\lor x_3)\land x_1$. The orange ellipses denote the function $(\neq_2)$.} 
	\label{sat}
	\end{figure}

	\begin{definition}
		A 3 conjunctive normal form (3CNF) formula $\phi$ on variables $x_1,x_2,...,x_n\in\{0,1\}$ is of the form $\bigwedge_i C_i$ where each \emph{clause} $C_i= \bigvee l_i^k$ with $k\leq3$ has each \emph{literal} $l_i^k=x_j$ or $l_i^k=\bar{x_j}$ for some $j\in[n]$.
		The problem \#3SAT is defined as:
		
		Input: A $3$-CNF formula $\phi$.
		
		Output: The number of satisfying assignments to $\phi$.
	\end{definition}
	
	\#3SAT is exactly a tensor network contraction problem parameterized by a set of Boolean symmetric functions $\mathcal{E}\cup\{OR_1,OR_2,OR_3,\neq_2\}$, where $\mathcal{E}=\{=_1,=_2,=_3,...\}$ denotes the set of all equality functions, $OR_1=[0,1]$, $OR_2=[0,1,1]$, and $OR_3=[0,1,1,1]$.
	For example, given a $3$-CNF formula $\phi=(x_1\lor x_2\lor \bar{x_3})\land (\bar{x_1}\lor x_3)\land x_1$, we can construct a tensor network $G$, showed on Figure \ref{sat}, with $Z(G)=$\#$3$SAT($\phi$).

\subsection{Tensor rank and decomposition}

	\quad\ A $k$-arity function or $k$-dimensional tensor $F \in \mathbb{C}^{q_1\times q_2\times \cdots \times q_k}$ is \emph{rank-one} if it can be written as an outer product of $k$ vectors or unary functions, i.e., $F=a^{(1)}\otimes a^{(2)} \otimes...\otimes a^{(k)}$, where $a^{(i)}$ is a vector for any $i\in [k]$. This means each element of $F$ is the product of corresponding vector elements, i.e., $F_{i_1i_2...i_k}=a^{(1)}_{i_1}a^{(2)}_{i_2}\cdots a^{(k)}_{i_k}$ for all $i_j\in [q_j]$. 
	The CANDECOMP/PARAFAC (CP) decomposition \cite{CPdecomposition} decomposes a $k$-dimensional tensor $F$ into a sum of component rank-one tensors, i.e., a CP decomposition of $F$ is $\sum_{r=1}^{R} a_r^{(1)}\otimes a_r^{(2)}\otimes \cdots \otimes a_r^{(k)}$ for some integer $R$.
	The \emph{rank} of $F$ is defined as 
	\begin{equation}
		rank(F)= \min \{R : F = \sum_{r=1}^{R} a_r^{(1)}\otimes a_r^{(2)}\otimes \cdots \otimes a_r^{(k)}\}.
	\end{equation}
	The expression $\sum_{r=1}^{rank(F)} a_r^{(1)}\otimes a_r^{(2)}\otimes \cdots \otimes a_r^{(k)}$ denotes a minimum CP decomposition of $F$.
	More basic information about tensor rank and decomposition can be seen in \cite{tensor-rank,CLOPZM2016}.
	
	A tensor $F\in \mathbb{C}^{q_1\times q_2\times \cdots \times q_k}$ is a symmetric tensor if all its dimensions are identical, i.e., $q_1=q_2=...=q_k$, and its elements are invariant under any permutation of the indices. For a symmetric tensor $F\in \mathbb{C}^{q\times q\times \cdots \times q}$, the \emph{symmetric rank} of $F$ is defined as:
	\begin{equation}
		rank_S(F)= min \{ R: F=\sum_{r=1}^{R} a_r\otimes a_r\otimes \cdots\otimes a_r=\sum_{r=1}^{R} a_r^{\otimes k}\}.
	\end{equation}
	Comon et al. show that $rank_S(F)\leq \binom{q+k-1}{k}$ in \cite{symmterictensor}.
	 
\subsection{Exponential time hypothesis}
	
	\quad\  Impagliazzo et al. \cite{ETHKSAT,SETH} introduced the \emph{Exponential Time Hypothesis} (ETH), which states SAT is not tractable in sub-exponential time.
	Dell et al. \cite{ETHpermTutte} put up the more relaxed counting version: \#ETH. 

	\begin{description}
		\item[\emph{\#ETH\cite{ETHpermTutte}}: ]\emph{
		There is a constant $\varepsilon>0$ such that no deterministic algorithm can compute \#3SAT in $2^{\varepsilon n}$ time, where $n$ is the number of variables of the input formula. }
	\end{description}

	The lower bound can be strengthened to $2^{\varepsilon m}$ according to the \emph{Sparsification Lemma} \cite{SETH}, where $m$ denotes the number of clauses in the input formula. 
	Liu \cite{exponentialdichotomyofCSP} proved such a sub-exponential time lower bound for the restriction to \#3SAT that every input $3$-CNF formula contains each variable in at most $3$ clauses. The restriction defines the problem \#$\{=_1,=_2,=_3,OR_2,OR_3,\neq_2\}$ where $(=_1)=[1,1]$ and $(=_3)=[1,0,0,1]$.
	It can be reduced to the problem \#$\{=_2,=_3,OR_3,\neq_2\}$ since we can simulate $=_1$ by $(=_1)(x)=\sum_{y,z\in\{0,1\}}(=_3)(x,y,z)(=_2)(y,z)$, $OR_1$ by $OR_1(x)=OR_3(x,x,x)$, and $OR_2$ by $OR_2(x,w)=\sum_{y,z\in\{0,1\}} OR_3(x,y,z)(=_3)(y,z,w)$, where $x,w\in\{0,1\}$. 
	
	\begin{lemma}\cite{exponentialdichotomyofCSP}
		There is a constant $\varepsilon>0$ such that the problem \#$\{=_2,=_3,\neq_2,OR_3\} $can not be computed in $2^{\varepsilon n}$ time if \#ETH holds, where $n$ is the number of vertices in the input.
		\label{lower_bound_of_3SAT}
	\end{lemma}

%% file: algorithm.tex
\section{Algorithms for contracting tensor networks}	
	\quad\ Kourtis et al. \cite{fastcounting} introduced an algorithm to contract a planar Boolean tensor network in $2^{O(\sqrt{\Delta}N)}$ time, where $N$ denotes the number of vertices and $\Delta$ denotes the maximum degree. 
	They presented a divide and conquer algorithm to find a sequence of edge separators to partition the network to $N$ isolated tensors, according to Lemma \ref{planar_edge_separator}.
	Then they contracted the isolated tensors in the reversed order of partitioning. 
	The algorithm guaranteed that each tensor appearing in the contraction process has $O(\sqrt{\Delta N})$ dimension so that the Boolean tensor network can be contracted in $2^{O(\sqrt{\Delta N})}$ time. 
	
	Inspired by the algorithm, we consider the edge separator of a finite element graph. 

	\begin{theorem}
		Let $G$ be a finite element graph with $N$ vertices and maximum degree $\Delta$.
		Suppose each element of $G$ has no more than $d$ boundary edges.
		A balanced edge separator $C$ of $G$ can be found in polynomial time, with $|C|=O(d\sqrt{\max\{\Delta,d\}N})$.
	\label{thm1}
	\end{theorem}

	\begin{proof}
	Let $G^*$ denotes the planar skeleton of $G$. 
	Suppose $G^*$ has $f$ faces $L_1,L_2,...,L_f$ with more than $3$ boundary edges.
	We construct a planar graph $G^{**}$ from $G^{*}$ by adding a new vertex $w_i$  inside each face $L_i$ and connecting it with all boundary vertices.
	 $W=\{w_1,w_2,...,w_f\}$. 
	 The planar graph $G^{**}$ has $N+f\leq 2N$ vertices and maximum degree $\Delta^{**}\leq \mathrm{max}\{\Delta,d\}$.
	By Lemma \ref{separator_for_finite_element_graph}, we find a balanced edge separator $C^{**}$ with $|C^{**}|=O(\sqrt{\Delta^{**}N})$ in $O(N)$ time. 
	 
	 Suppose $C^{**}$ partitions $G^{**}$ into two disconnected parts $A^{**}$ and $B^{**}$.
	Let $A=A^{**}\cap V(G)$ and $B=B^{**}\cap V(G)$. 
	For an edge $(u,v)\in E(G)$ which connects $A$ and $B$, either $(u,v)\in C^{**}$ or $\{(u,w_i),(w_i,v)\}\cap C^{**}\neq\emptyset$ for some $w_i\in W$.
	If $(u,w_i)\in C^{**}$, we add all diagonals, which connect $u$ with some boundary vertex of $L_i$, to the set $C$. 
	We do the same if $(w_i,v)\in C^{**}$.
	Finally, we add $C^{**}\cap E(G)$ to $C$. 
	$C$ is a balanced edge separator of $G$ and $|C|=O(d\sqrt{\max\{\Delta,d\}N})$. 
	\end{proof}

	Then we can build an exponential algorithm similar to the algorithm in \cite{fastcounting}.

	\begin{theorem}
			Let $G$ be a finite element with $N$ vertices and maximum degree $\Delta$.
			Suppose each element of $G$ has no more than $d$ boundary edges.
			Given a Boolean tensor network whose underlying graph is $G$, it can be contracted in $2^{O(d\sqrt{\max\{\Delta,d\}N})}$ time.
	\label{thm2}
	\end{theorem}
	
\subsection{Defined on a set of Boolean symmetric functions}		
	
	\quad\  We consider accelerating the above algorithms. 
	When the tensor network is defined on a set of Boolean symmetric functions, we replace each function of some arity $n$ by a planar bounded degree gadget with $O(n)$ vertices.
	
\begin{figure}[ht]
		\centering
		\includegraphics[scale=0.25]{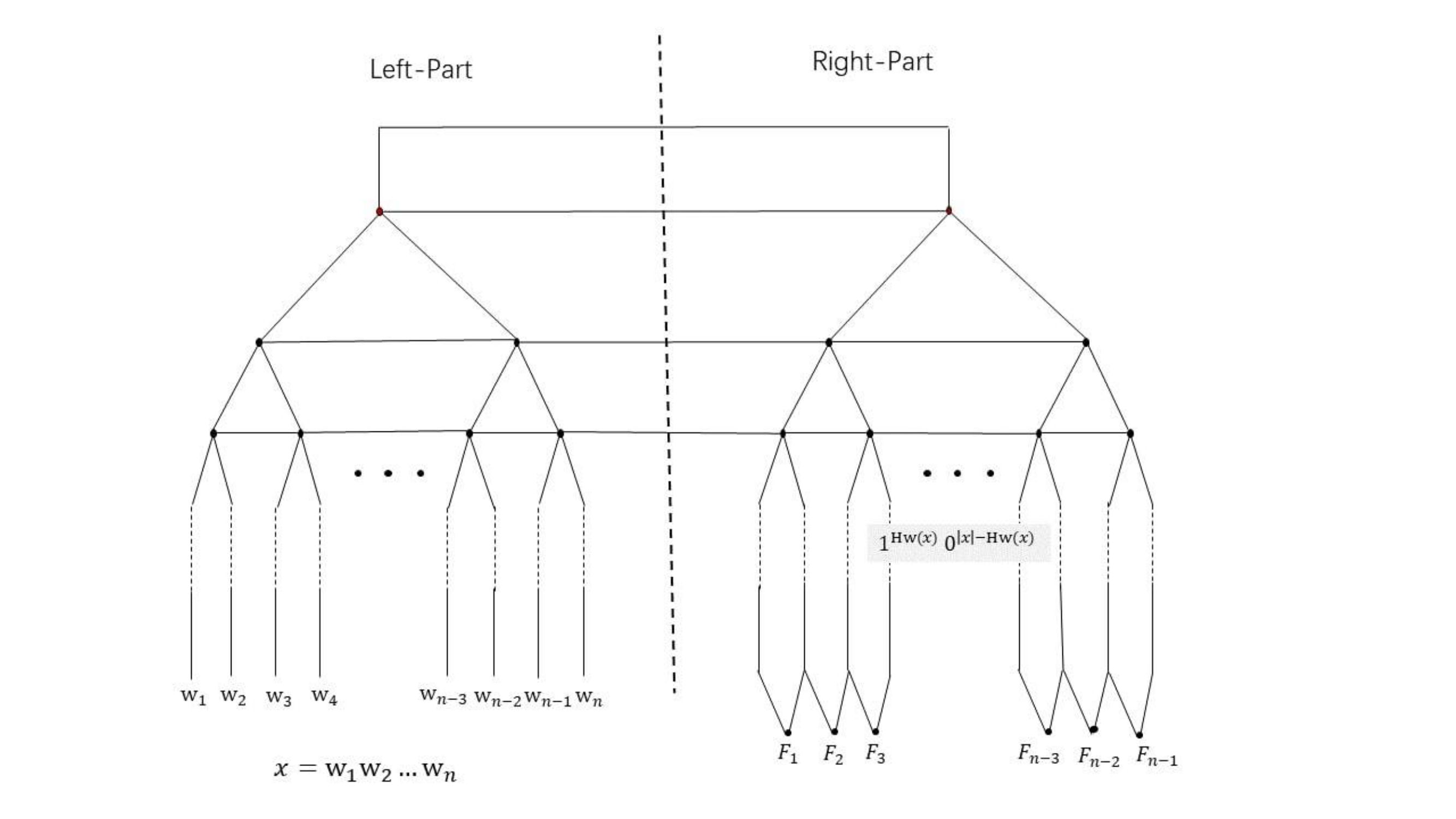}   
		\caption{ Planar bounded degree structure to realize all Boolean symmetric functions. } 
		\label{alg-2}
	\end{figure}

	Suppose $F=[f_0,f_1,\cdots,f_n]$ with $f_0,f_1,...,f_n\in\mathbb{C}$ (w.l.o.g, $n$ is a power of $2$)\footnote{We can use some additional edges, whose other endpoints are attached with the unary function $[1,0]$, to refill the arity. This operator only increases the aimed gadget size to double.}. We replace $F$ with an equivalent planar gadget, shown in Figure \ref{alg-2}. 
	The general idea of the gadget is to rearrange the assignment of variables since the order of elements in the assignment is irrelevant. Treating each assignment as an $n$-length string over $\{0,1\}$, we use the left part to count the number of $1$ (Hamming weight) and use the right part to return an ordered $n$-length string where all $1$ are in front of $0$. 
	Then we decide the corresponding function value according to the location of the border of $1$ and $0$. 
	Next, we introduce the gadget in detail.
	
\begin{figure}
		\begin{subfigure}[t]{0.5\textwidth}
			\centering
			\includegraphics[width=0.4\textwidth]{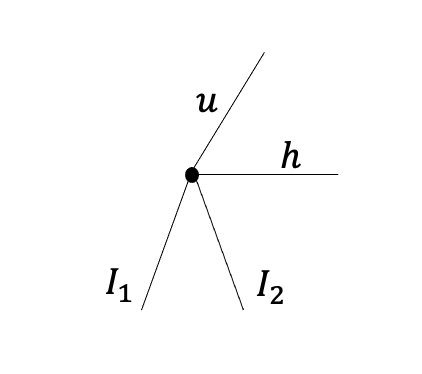}
			\includegraphics[width=0.4\textwidth]{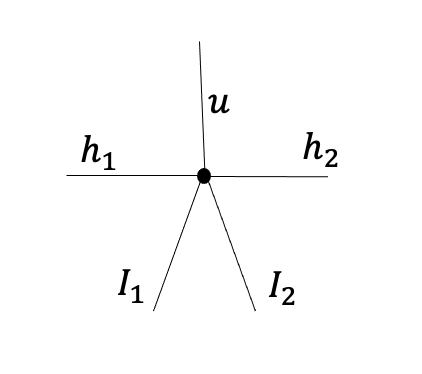}
			\caption{In the left part}
		\end{subfigure}\hfill
		\begin{subfigure}[t]{0.5\textwidth}
			\centering
			\includegraphics[width=0.4\textwidth]{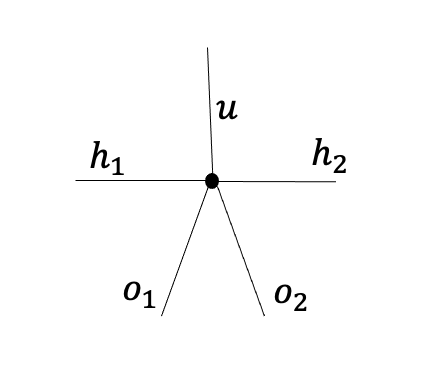}
			\includegraphics[width=0.4\textwidth]{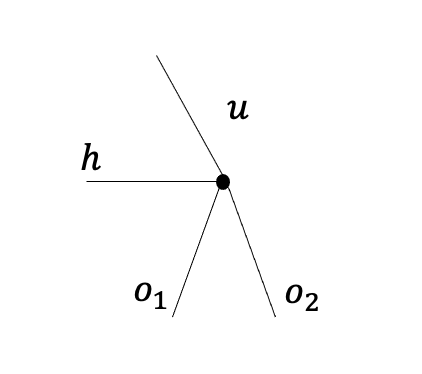}
			\caption{In the right part}
		\end{subfigure}
		\caption{Vertex modes of the planar gadget.}
		\label{vertexmode}
	\end{figure}

The left part of this planar gadget uses the idea ``Adder'' to calculate the binary expression of Hamming weight $Hw(x)$ of the assignment $x=(w_1w_2\cdots w_n)\in\{0,1\}^n$. The functions all are simple addition operators in the left part.
The left structure accepts $x$ and adds every two adjacent bits. 
Each vertex denotes an addition function $A$ or $B$, shown in Figure \ref{vertexmode}-(a), which adds two bits $I_1, I_2$ or three bits $h_1, I_1, I_2$, sets the most significant bit $u$ to join a higher level addition, and sets the least significant bit $h$ or $h_2$ to join the operation of the horizontal adjacent vertex on the right. 
After $\log_2 n$ levels, the left part outputs the binary expression of ${\rm Hw}(x)$ in the horizontal edges from top to bottom (information would not be lost since ${Hw}(x)\leq \log_2 n +1$).
	
	\begin{align*}
		&A(u,h,I_1,I_2)=
		\begin{cases}
			1  &  \text{if $u=\lfloor (I_1 + I_2)/2 \rfloor $ and $h=(I_1+I_2)$ mod 2} \\
			0 &  \text{else}\\
		\end{cases}
		\\
		&B(u,h_1,h_2,I_1,I_2)=
		\begin{cases}
			1  &  \text{if  $u=\lfloor(I_1 + I_2 + h_1)/2\rfloor$ and $h_2=(I_1+I_2+h_1)$ mod 2} \\
			0 &  \text{else}\\
		\end{cases}
	\end{align*}
	
	The right part uses the horizontal $(\log_2 n+1)$ bits to recover an ordered string of the form $1^{{\rm Hw}(x)} 0^{|x|-{\rm Hw}(x)}$ before outputting the accuracy value of $F$. It is obvious that the top two bits would not be $1$ at the same time. 
	In the right part, the functions are a little different from those in the left part, shown in Figure \ref{vertexmode}-(b). Each of them is one of the following functions:
	
	\begin{align*}
		&C(u,h,o_1,o_2)=
		\begin{cases}
			1 &  \text{if $o_1=u + h$ and $o_2= 2u+h-o_1$} \\
			0 &  \text{else}\\
		\end{cases}
		\\
		&D(u,h_1,h_2,o_1,o_2)=
		\begin{cases}
			1 &  \text{if $u=1$ \& $o_1=o_2=1$ and $h_2=h_1$} \\
			1 &  \text{if $u=0$ \& $o_1=h_1$ and $o_2=h_2=0$} \\
			0 &  \text{else}\\
		\end{cases}
	\end{align*}
	
	The Hamming weight is reflected by the location of the sub-string $10$
	in the ordered string $1^{{\rm Hw}(x)} 0^{|x|-{\rm Hw}(x)}$. The gadget uses additional $2$-arity functions $F_1,\cdots,F_i,\cdots,F_{n-1}$ to identify the location, where $i\in \{2,\cdots,n-2\}$.

	\begin{equation}
	\begin{cases}
	F_1(0, 0)=f_0, &   \\
	F_1(1, 0)=f_1, &  \\
	F_1(1, 1)=F_1(0,1)=1; &  
	\end{cases}
	\nonumber
	\end{equation}
	
	\begin{align*}
		&\begin{cases}
			F_i(0,0)=F_i(1,1)=F_i(0,1)=1, &   \\
			F_i(1,0)=f_i; &  
		\end{cases}
	\\
		&\begin{cases}
			F_{n-1}(1,1)=f_n, &   \\
			F_{n-1}(1,0)=f_{n-1}, &  \\
			F_{n-1}(0,0)=F_{n-1}(0,1)=1. &  
		\end{cases}
	\end{align*}
	The number of vertices in such a planar gadget is $2(n+ \frac{n}{2} + \frac{n}{4} + \cdots +1 ) + n-1 = O(n)$, and the maximum degree is $5$. 
	
	For any tensor network $G$ defined on the set of Boolean symmetric functions, we preprocess it to a bounded degree tensor network $G'$ by the above gadgets.
	If $G$ is planar, then $\sum_{v\in V(G)} d_{v}=2E(G)\leq 6|V(G)|-12$. So
	$G'$ has $O(|V(G)|)$ vertices. We apply the algorithm \cite{fastcounting} on $G'$. 
	
	\begin{theorem}
		Any planar tensor network consisting of $N$ Boolean symmetric tensors can be contracted in $2^{O(\sqrt{N})}$ time. 
		\label{thm4}
	\end{theorem}
	
\begin{figure}[ht]
	\centering
	\includegraphics[scale=0.4]{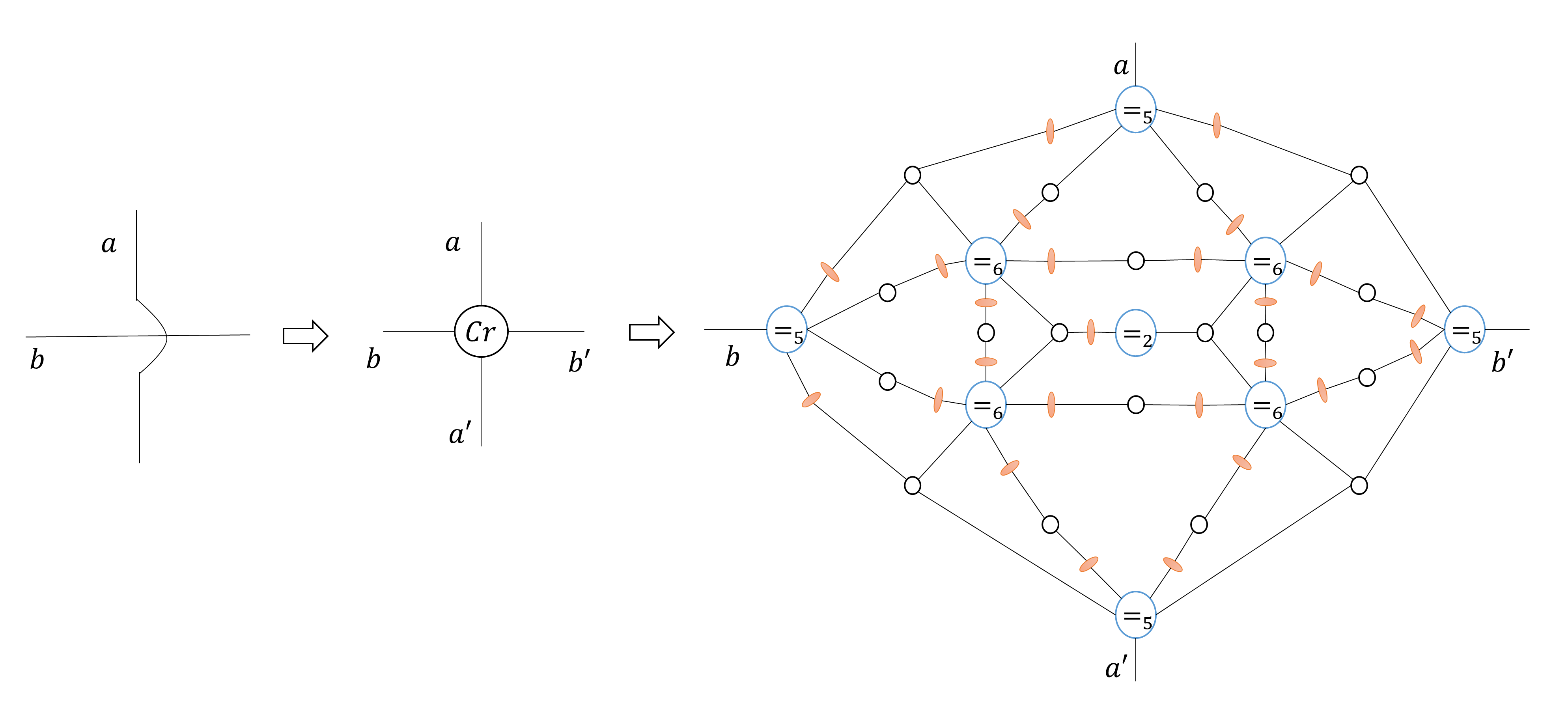}   
	\caption{ A planar gadget for crossing. The black circle vertices are assigned $OR_2$ or $OR_3$; the orange ellipses are the vertices assigned $(\neq_2)$.} 
	\label{crossing}
	\end{figure}

	If $G$ is a finite element graph, we need more steps to preprocess $G$. 
	We transfer $G$ to a planar graph first. Suppose there are $f$ elements $L_1,L_2,...,L_f$ with more than $3$ boundary edges in the planar skeleton $G^*$ of $G$. $d_i$ denotes the number of boundary edges in $L_i$ for $i\in[f]$.    
	There are no more than $(\sum_{i\in [f]} d_i^4)$ crossings in $G$, according to the definition of finite element graphs. 
	We replace each crossing with a new vertex assigned with the function $Cr$, shown in Figure \ref{crossing}. The function $Cr$ keeps $a=a'$ and $b=b'$ for Boolean variables $a,a',b,b'$.

	Then we obtain a planar tensor network $G'$with no more than $N+\sum_{i\in [f]} d_i^4\leq N+2m d^3\leq N+ 6Nd^3$ vertices, where $m=|E(G^*)|\leq |E(G)|$ and $d=\mathrm{max}\{d_1,...,d_f\}$. $Z(G')=Z(G)$.
	A gadget with the function $Cr$, shown in Figure \ref{crossing}, consists of only Boolean symmetric functions. 
	We construct the planar $G''$ from $G'$ by replacing each occurrence of $Cr$ with such a gadget.
	$G''$ has $O(d^3N)$ vertices and maximum degree no more than $6$. 
	$Z(G'')=Z(G')=Z(G)$.
	We can compute $Z(G'')$ in $2^{O(\sqrt{d^3N})}$ time by Theorem \ref{thm4}. 
	Since $G''$ is constructed in polynomial time, the above algorithm computes $Z(G)$ in $2^{O(\sqrt{d^3N})}$ time.
	\begin{theorem}
		A tensor network consisting of $N$ Boolean symmetric tensors can be contracted in $2^{O(\sqrt{d^3N})}$ time if the underlying graph is a finite element graph whose elements all have no more than $d$ boundary edges.
		\label{thm6}
	\end{theorem}
	
	Can we also construct a planar bounded degree gadget for any symmetric function over a larger domain, for example, the domain $[3]$? The algorithm in Theorem \ref{thm3} can be extended further if we can. Unfortunately,  the answer is negative, according to Appendix A.
		
	\subsection{Defined on a set of finite functions}
	
		\quad\ It is trivial that a tensor network consisting of only unary functions can be contracted in polynomial time. $CP$ decomposition provides the way to decompose a tensor to a series of unary functions. 
		
		\begin{theorem}
			Let $\mathcal{F}$ be a finite set of functions and $R=\mathrm{max}\{rank(F)|F\in\mathcal{F}\}$.
			A planar tensor network defined on $\mathcal{F}$ can be contracted in $R^{O(\sqrt{N})}$ time, where $N$ denotes the number of vertices in the input.
			\label{thm3}
		\end{theorem}
		
		\begin{proof}
			We state the main idea of the divide and conquer algorithm here. 
			Appendix A.1 shows details.
			
			Given a planar tensor network $G$ with $N$ vertices, we search for a balanced node separator $C$ with $|C|=O(\sqrt{N})$ in linear time, by Lemma \ref{planar_node_separator}. 
			For each vertex $v\in C$, we replace the $d_v$-dimensional tensor $F_v$ by the components of a minimum $CP$ decomposition of $F_v$ independently. Suppose $F_v=\sum_{i=1}^{r} u_{i}^{1}\otimes u_{i}^{2}\otimes \dots \otimes  u_{i}^{d_v}$, where  $r=rank(F_v)\leq R$, then we obtain a series of new tensor networks $G_1,...,G_r$ by  replacing $F_v$ with $r$ components $u_{1}^{1}\otimes u_{1}^{2}\otimes \dots \otimes  u_{1}^{d_v},...,u_{r}^{1}\otimes u_{r}^{2}\otimes \dots \otimes  u_{r}^{d_v}$ independently.
			We make a contraction between each $u_{i}^{j}$ and its adjacent tensor in $G_i$, where $j\in[k]$. 
			After the contractions, $G_i$ is a tensor network consisting of two disconnected planar tensor networks $A_i$ and $B_i$, where each has no more than $\frac{2}{3}N$ vertices. 
			Suppose $Z(G)$ denotes the value of $G$. $Z(G)=\sum_{i=1}^{r} Z(G_i)=\sum_{i=1}^{r}  Z(A_i)Z(B_i)$. 
			Then we compute $Z(A_i)$ and $Z(B_i)$ for $i\in[r]$.
			
			The value of $G$ can be computed in $R^{O(\sqrt{N})}$ time by the above algorithm. The runtime analysis is presented in Appendix B.
 		\end{proof}
 	
		The above algorithm can be extended for contracting tensor networks on finite element graphs, by Lemma \ref{separator_for_finite_element_graph}. 
		
		\begin{theorem}
			Let $\mathcal{F}$ be a finite set of functions and $R=\mathrm{max}\{rank(F)|F\in\mathcal{F}\}$.
			A tensor network defined by $\mathcal{F}$, whose underlying graph is a finite element graph having no elements with more than $d$ boundary edges,
			can be contracted in $R^{O(d\sqrt{N})}$ time, where $N$ is the number of tensors.
			\label{thm5}
		\end{theorem}

%% file: lowerbound.tex
\section{Lower bounds of tensor network contraction problems}

	\quad\ In this section, we prove the lower bound for contracting tensor networks, even restricting the underlying graphs to planar graphs or finite element graphs. 
	
	\begin{theorem}
		If \#ETH holds, then there is a constant $\varepsilon>0$ such that a planar tensor network can not be contracted in $2^{\varepsilon \sqrt{N}}$ time, where $N$ denotes the number of vertices in the input.
		
		Furthermore, the result holds for the planar tensor networks defined by the set $\{=_2,=_3,\neq_2,OR_3\}$. 
		\label{thm7}
	\end{theorem}
	
	\begin{proof}
		We reduce the problem \#$\{=_2,=_3,\neq_2,OR_3\}$ to $pl$-\#$\{=_2,=_3,\neq_2,OR_3\}$. 
		Let $G$ with $N$ vertices be an instance of \#$\{=_2,=_3,\neq_2,OR_3\}$.
		$G$ has at most $3N$ edges and $9N^2$ crossings. 
		
		We replace each crossing with a new vertex assigned the function $Cr$. 
		The new tensor network $G'$ is an instance of $pl-\#\{=_2,=_3,\neq_2,OR_3,Cr\}$.  
		$G'$ has at most $(N+9N^2)$ vertices. 
		We replace each occurrence of $Cr$ with the gadget shown in Figure \ref{crossing}, then we construct
		a planar tensor network $G''$with $O(N^2)$ vertices. 
		$G''$ is an instance of $pl$-\#$\{=_2,=_3,\neq_2,OR_3,OR_2,=_5,=_6\}$. 
		We further replace each occurrence of $OR_2$, $=_5$, or $=_6$ by the gadgets shown in Figure \ref{gadgets}.
		The generated tensor network $G'''$ is an instance of $pl$-$\#\{=_2,=_3,\neq_2,OR_3\}$. $G'''$ has $O(N^2)$ vertices.
		$Z(G''')=Z(G'')=Z(G')=Z(G)$.
		\begin{figure}
			\begin{subfigure}[t]{0.33\textwidth}
				\centering
				\includegraphics[width=0.7\textwidth]{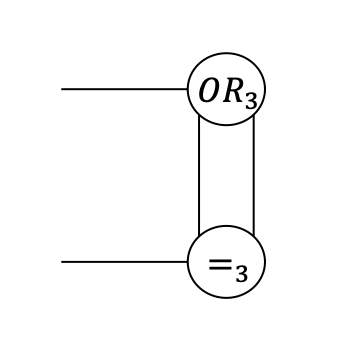}
				\caption{with the function $OR_2$.}
			\end{subfigure}\hfill
			\begin{subfigure}[t]{0.33\textwidth}
				\centering
				\includegraphics[width=0.6\textwidth]{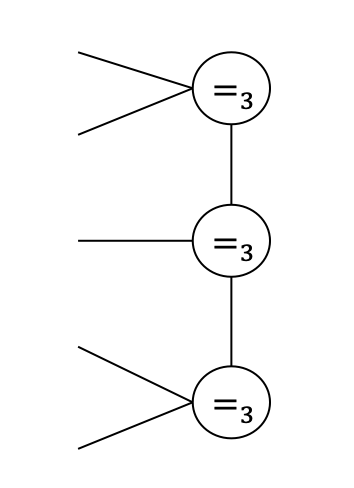}
				\caption{with the function $=_5$.}
			\end{subfigure}
			\begin{subfigure}[t]{0.33\textwidth}
			\centering
			\includegraphics[width=0.55\textwidth]{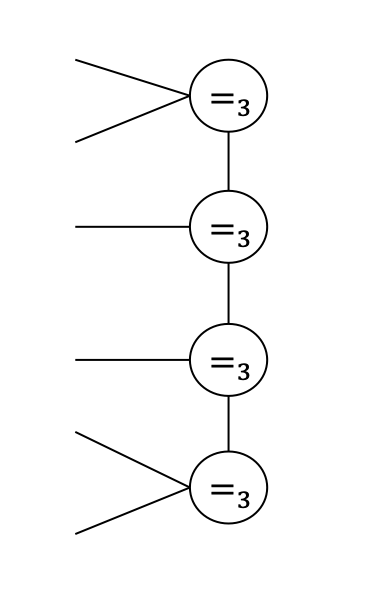}
			\caption{with the function $=_6$.}
			\end{subfigure}
			\caption{Some $\{OR_3,=_3\}$-gates.}
			\label{gadgets}
		\end{figure}
	
	Suppose the theorem is false, i.e., we can solve $Z(G''')$ in $2^{\varepsilon \sqrt{cN^2}}$ time for any $\varepsilon>0$, then we can solve $Z(G)$ in $\mathrm{poly}(N)+2^{\varepsilon \sqrt{cN^2}}\leq2^{\varepsilon'N}$ time for some constant $\varepsilon'$. It is a contradiction to Lemma \ref{lower_bound_of_3SAT}. 
	\end{proof}

	Now we think about the lower bound for tensor network contraction problems on finite element graphs. Given a planar graph, we use triangular partitioning to build a finite element graph. 
	
	\begin{theorem}
		If \#ETH holds, then there is a constant $\varepsilon>0$ such that a tensor network, whose underlying graph is a finite element graph with $N$ vertices, can not be contracted in $2^{\varepsilon \sqrt{N}}$ time.
		
		Furthermore, the result holds even when the parameterized set is restricted to Boolean symmetric functions of arity no more than $7$.
		\label{thm8}
	\end{theorem}
	
	\begin{proof}
		Let $G$ be an instance of $pl-\#\{=_2, =_3, \neq_2, OR_3\}$. The underlying graph has a planar embedding, which $ G$ also denotes. 
		For the embedded planar graph $G$, we think about triangular partitioning every face with more than $3$ boundary edges.
		We first deal with \emph{bridges}. We add a new vertex $w$ and edges $(v_1,w),(v_2,w)$ for each bridge $(v_1,v_2)$. 
		Then we get a new planar graph $G'$ with no bridge. Let $L_1,L_2,...,L_s$ denote the faces with more than $3$ boundary edges in $G'$. 

		Suppose the boundary vertices of a face $L\in\{L_1,L_2,...,L_s\}$ are labeled $v_1,v_2,...,v_{d}$ in clockwise order \footnote{A vertex is given different labels if more than one of its incident edges are in the boundary of $L$.}.
		We add a $(\lceil \frac{d}{2}\rceil)$-length cycle $(u_1,u_2,...,u_{\lceil d/2 \rceil },u_1)$ inside $L$, and add edges $(u_j,v_{2j-1}),(u_j,v_{2j}),(u_j,v_{2j+1})$ for $j\in\{1,2,..,\lfloor d/2 \rfloor\}$, where $v_{d+1}$ is exact $v_1$.
		If $d$ is odd, we extra add the edges $(u_{\lceil d/2 \rceil},v_{d})$ and $(u_{\lceil d/2 \rceil},v_{d_1})$. 
		We partition $L$ into some triangles and a face $L'$ with $(\lceil d/2 \rceil )$ boundary edges. 
		We continue to partition $L'$ in the same way. 
		After no more than $\lceil log_2 d \rceil$ rounds, we completely triangulate $L$. 
		We add no more than $d$ new vertices with degree no more than $7$. 
		For example, we show in Figure \ref{triangulation} the process of partitioning a face with $7$ boundary edges.
		\begin{figure}[ht]
			\centering
			\includegraphics[scale=0.2]{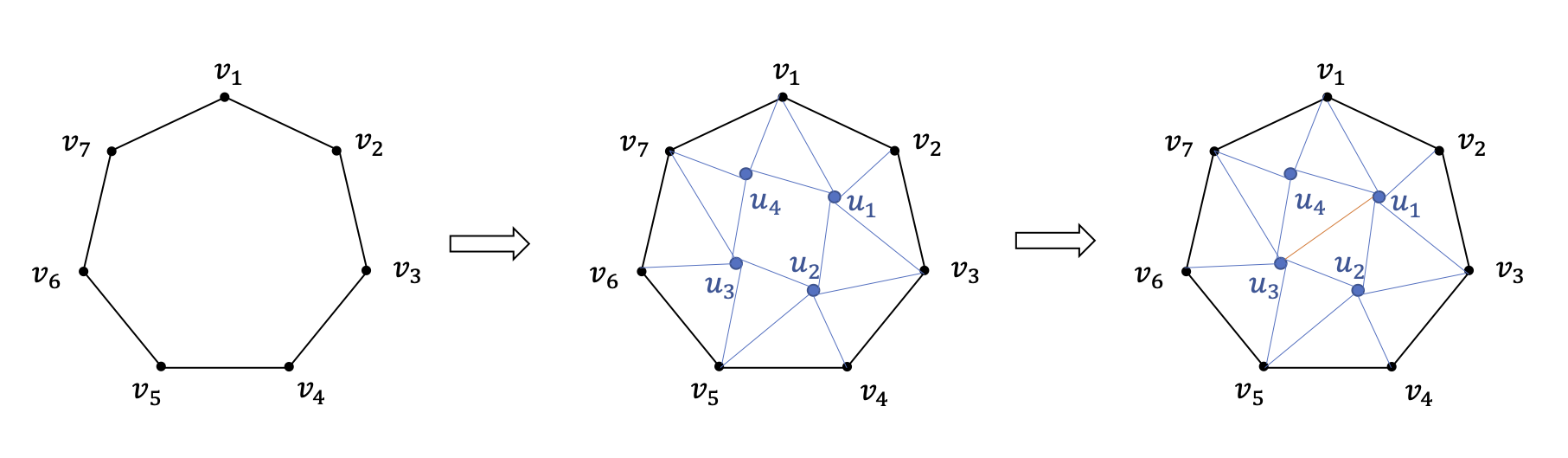}   
			\caption{ The process of triangular partitioning a face with $7$ boundary edges. } 
			\label{triangulation}
		\end{figure} 
		
		We triangulate each face in $G'$ and obtain a finite element graph $G''$ in $\mathrm{poly}(N)$ time. 
		The planar skeleton of $G''$ is itself.  
		$G''$ has no more than $N+\sum_{L_i} d_i\leq N+2|E(G')|\leq6N$ vertices, where $d_i$ is the number of boundary edges in face $L_i$. 
		The maximum degree of $G''$ is no more than $7$. 
		
		Next, we assign to every vertex $v\in V(G'')$ an appropriate Boolean symmetric function, such that $Z(G')=Z(G)$.		
		For each vertex $v\in V(G'') - V(G)$, we assign the function $[1,0]^{\otimes d_v}$ to $v$, so that the edge-variables in $E(G'')-E(G)$ must be assigned $0$.
		For each vertex $v\in V(G'')\cap V(G)$, we assign to $v$ the function $F_v$, according to the function $[f_0,f_1,f_2,f_3]$ (or $[f_0,f_1,f_2]$) which is originally assigned to $v$ in $G$. 
		$F_v$ denotes the function $[0,...,0,f_0,f_1,f_2,f_3]$ (or function $[0,...,0,f_0,f_1,f_2]$), which takes value $0$ when the Hamming weight of input is no more than $k-4$ (or $k-3$). 
		The assignment to edges in $E(G'')-E(G)$ are fixed and make no difference to the value of $G''$.
		The assignments to edges $E(G’')\cup E(G)$ contribute to $Z(G'')$ in the same way as they contribute to $Z(G)$.  So $Z(G‘')=Z(G)$.
		
		Suppose $Z(G'')$ can be solved in $2^{\varepsilon\sqrt{6N}}$ time for any $\varepsilon>0$, then we have an algorithm to compute $Z(G)$ in $(\mathrm{poly}(N)+2^{\varepsilon\sqrt{6N}})\leq2^{\varepsilon’ \sqrt{N}}$ time for some $\varepsilon'>0$. It is a contradiction to Theorem \ref{thm7}.  
	\end{proof}

%% file: conclusion.tex
\section{conclusion}

	\quad\ In this article, we introduce some efficient algorithms for tensor network contraction problems on two special graph structures: planar graphs and finite element graphs. We put up different methods to design the algorithms, depending on the classes of the parameterized sets.
	
	When the parameterized sets are restricted to Boolean symmetric functions, we also utilize \#ETH to prove a tight lower bound for tensor network contraction problems on the two special graph structures.
	
	There are still gaps between the algorithms and the lower bounds.
	

%% file: appendix.tex
\section{Planar bounded degree gadgets for symmetric functions over a large domain} 

		\quad\  Utilizing the planar gadget of crossing in Figure \ref{crossing}, we can relax the planar restriction to the gadget. 
We want to build a $O(n)$-size bounded degree gadget with only $O(n)$ crossings for any $n$-arity symmetric functions over the domain $[q]$ with $q \geq 3$.
The answer is negative. We prove this by the special case: $q=3$. 

There is $2^{\frac{(n+1)(n+2)}{2}}$ different symmetric functions which map  $\{0,1,2\}^n$ to $\{0,1\}$. According to the lower bound of Kolmogorov complexity (Thm 14.2.4 in \cite{impressed}),
the core-word to encode such a function is at least $\frac{(n+1)(n+2)}{2}$ bits in length.

Consider encoding a $c\cdot n$-size gadget $G$ with maximum degree $\Delta$, where $c$ and $\Delta$ are some constants. We use the function $C(\Delta)$ to denote the number of functions of arity no more than $\Delta$. Since $\Delta$ is a constant, $C(\Delta)$ is also a constant. For each vertex in $G$, we can use a $log_2 (cn)+log_2 C(\Delta)$-length core-word to encode it, where the first $log_2 (cn)$ bits encode the index of the vertex and the next $log_2 C(\Delta)$ bits indicate the function assigned to the vertex. 
Each edge $G$ can be encoded as a pair of core-words of two adjacent nodes. Then the gadget has a core-word, whose length is no longer than $\Delta cn\cdot2(log_2 (cn)+log_2 C(\Delta))$ bits. 
It is impossible to compress any $n$-arity symmetric function, defined on $\{0,1,2\}$, into a $\Delta cn\cdot2(log_2 (cn)+log_2 C(\Delta))$ length core-word without information loss, when $n$ is big enough. 

Finding a planar bounded degree gadget for every symmetric function is impossible. 
Nevertheless, we can construct such gadgets for some special classes of functions. One class is $\{F: [\{0,1,\cdots,q\}]^n\to  \mathbb{C}$ $|$ $F(x_1,x_2,\cdots,x_n)=F_1(x_1,x_2,\cdots,x_n)\otimes F_2(x_1,x_2,\cdots,x_n)\otimes\cdots \otimes F_q(x_1,x_2,\cdots,x_n), n\in \mathbb{N}\}$, where the value of $F_i$ is only decided by the number of $i$ in the assignment of variables. We can still simulate $F$ by counting the numbers of $i$ and computing each $F_i$ independently, similar to the one in Figure \ref{alg-2}. The left part computes $q$ groups of binary expressions, which indicate the number of $1,2,\cdots,q$ in the assignment. 
The degree of a vertex in the left part increases to $2q$ or $3q-1$. 
There are $q$ copies of the right part in Figure \ref{alg-2}, where the $i$-th duplicate recovers $F_i$ independently. The copies would bring $O(q^2n)$ crossings.
The gadget has only $O(qn)$ nodes and $O(q^2n)$ crossings in total, and the maximum degree is $O(q)$. 

\section{The algorithm to contract planar tensor networks in $\mathbf{R^{O(\sqrt{N})}}$ time}

	\quad\ Let $\mathcal{F}$ be a finite set of functions and $R=\max\{rank(F) \mid F\in\mathcal{F}\}$.
	We compute and record a minimum CP decomposition for every function in $\mathcal{F}$. 
	$TD(\mathcal{F})$ record all unary functions appearing in the decomposition. 
	We further compute the set $\langle\mathcal{F}\rangle=\mathcal{F} \cup \{ F \circ u \mid F\in \langle\mathcal{F}\rangle$ and $u\in TD(\langle\mathcal{F}\rangle)\}$, and record the minimum CP decomposition of each function in $\langle\mathcal{F}\rangle$. 
	$F\circ u$ denotes any $(k-1)$-arity function in $\{$ $\sum_{x} F\cdot u(x)\mid$ $x$ is a variable of $F$ $\}$, where $F$ is a function of some arity $k$ and $u$ is a unary function. 
	Above preparation aims to determine all possible functions appearing in the algorithm process and record their minimum CP decompositions.  
	Since $\mathcal{F}$ are finite, we can complete the preparation in finite time.

	The algorithm works as follows.

\begin{algorithm}[h]
	\caption{Contracting a planar tensor network defined on $\mathcal{F}$}
	\LinesNumbered
	\KwIn{$G$.}
	\KwOut{$Z(G)$.}
	
	\eIf{$|V|>1$}{
		Find a node separator $S$ of $G$.
		
		\ForEach{$v \in S$}{
			\If{$F_v=\sum_{i=1}^{rank(F_v)} u_{i}^{1}\otimes u_{i}^{2}\otimes \dots \otimes  u_{i}^{d_v}$} {
				Replace $F_v$ by $\{u_{1}^{1}\otimes\dots\otimes u_{1}^{d_v}\}$, $\dots$, $\{u_{rank(F_v)}^{1}\otimes\dots\otimes u_{rank(F_v)}^{d_v}\}$ respectively.
			}
		}
	
		Rename every new tensor network $G_{l_1,l_2,\cdots,l_{|S|}}$, where $l_j\in \{1,2,\cdots,R\}$ for $j={1,2,\cdots,|S|}$. 
		
		\ForEach{$G_{l_1,l_2,\cdots,l_{|S|}}$}{
			Adjust $G_{l_1,l_2,\cdots,l_{|S|}}$ by contracting every unary function, which comes from the functions in $S$.
			Then $G_{l_1,l_2,\cdots,l_{|S|}}$ is divided to two disconnected components $A_{l_1,l_2,\cdots,l_{|S|}}$ and $B_{l_1,l_2,\cdots,l_{|S|}}$. \\
			Solve $Z(A_{l_1,l_2,\cdots,l_{|S|}})$ by Algorithm 1.\\
			Solve $Z(B_{l_1,l_2,\cdots,l_{|S|}})$ by Algorithm 1. \\
			$Z(G_{l_1,l_2,\cdots,l_{|S|}})= Z(A_{l_1,l_2,\cdots,l_{|S|}})Z(B_{l_1,l_2,\cdots,l_{|S|}})$.
		}
		
		$Z(G)=\sum_{l_1,l_2,\cdots,l_{|S|}} Z(G_{l_1,l_2,\cdots,l_{|S|}})$.
	}{
		Compute $Z(G)$ directly.
	}
	
	\Return $Z(G)$.
	
\end{algorithm}

	Consider the time recursive formula of Algorithm 1. 
	Suppose the input tensor network $G$ defined on $\mathcal{F}$ has $N$ vertices.
	We can find a balanced node separator $S\in V(G)$ with $|S|\leq 2\sqrt{2N}$ in $c_1N$ time for some constant $c_1$, by Lemma \ref{planar_node_separator}. $S$ partitions $G$ into two disconnected parts where each has no more than $\frac{2}{3}N$ vertices. 
	We independently replace the functions in $S$ with the corresponding components in their minimum $CP$ decompositions. 
	Since we have pre-recorded all possible minimum $CP$ decompositions and the results of the possible contractions in the adjustment stage, every new tensor network $G_{l_1,l_2,\cdots,l_{|S|}}$ can be constructed and adjusted in $\mathrm{poly}(N+\Delta |S|)$ time. 
	The algorithm generates at most $R^{|S|}$ new tensor networks, each of which consists of two disconnected sub-networks $A_{l_1,l_2,\cdots,l_{|S|}}$ and $B_{l_1,l_2,\cdots,l_{|S|}}$ with $|V(A_{l_1,l_2,\cdots,l_{|S|}})|,|V(B_{l_1,l_2,\cdots,l_{|S|}})|\leq \frac{2}{3}N$.
	
	If we have the values $Z(A_{l_1,l_2,\cdots,l_{|S|}})$ and $Z(B_{l_1,l_2,\cdots,l_{|S|}})$, we compute the value of $G_{l_1,l_2,\cdots,l_{|S|}}$ by a multiplication. Furthermore, we obtain $Z(G)$ by $R^{|S|}$ times additions after obtaining all values of the new graphs. 
	
	The time recursive formula is:
	\begin{equation}
		T(N)= c_1\cdot N + R^{|S|}(\mathrm{poly}(N+ \Delta |S|)+( T(aN) + T(bN) ) )+ \mathrm{poly}(R^{|S|}), \quad\ a,b\leq\frac{2}{3}.
		\label{equ-1}
		\nonumber
	\end{equation}
	
	By mathematical induction, the total time of Algorithm 1 is $R^{O(\sqrt{N})}$.